\documentclass[aps,pre,twocolumn,footinbib,superscriptaddress]{revtex4-1}
\usepackage[T2A]{fontenc}
\usepackage[utf8]{inputenc}
\usepackage{amsmath}
\usepackage{amssymb}
\usepackage{color,xcolor}
\usepackage{natbib}
\usepackage[colorlinks=true,bookmarks=false,citecolor=blue,urlcolor=blue]{hyperref}
\usepackage{bm}
\usepackage{epsfig}
\usepackage{verbatim} 
\usepackage{morefloats}
\usepackage{graphicx}
\usepackage{bibentry}
\usepackage[mathscr]{euscript}
\usepackage{enumerate}
\usepackage{ bbold } 

\usepackage{subfigure}
\usepackage{amsmath}
\usepackage{amssymb}

\usepackage{ amssymb }
\usepackage{bm}

\def\dfrac{\displaystyle\frac}  

 \usepackage{blindtext}
 
\renewcommand{\phi}{\varphi}

\usepackage{hyperref}

\begin{document}

\title{Data-driven model reconstruction for nonlinear wave dynamics}

\author{Ekaterina Smolina}
\affiliation{Department of Control Theory, Nizhny Novgorod State University, Gagarin Av. 23, Nizhny Novgorod, 603950 Russia}
\author{Lev Smirnov}
\affiliation{Department of Control Theory, Nizhny Novgorod State University, Gagarin Av. 23, Nizhny Novgorod, 603950 Russia}
\author{Daniel Leykam} 
\affiliation{Science, Mathematics and Technology Cluster,\\ Singapore University
of Technology and Design, 8 Somapah Road, 487372 Singapore}
\author{Franco Nori}
\affiliation{Theoretical Quantum Physics Laboratory, Cluster for Pioneering Research, RIKEN, Wakoshi, Saitama 351-0198, Japan}
\affiliation{Center for Quantum Computing (RQC), RIKEN, Wako-shi, Saitama 351-0198, Japan}
\affiliation{Physics Department, University of Michigan, Ann Arbor, MI 48109-1040, USA}
\author{Daria Smirnova}
\affiliation{Theoretical Quantum Physics Laboratory, Cluster for Pioneering Research, RIKEN, Wakoshi, Saitama 351-0198, Japan}
\affiliation{Research School of Physics, Australian National University, Canberra, ACT 2601, Australia}

\begin{abstract}
The use of machine learning to predict 
wave dynamics is a topic of growing interest, but commonly-used deep learning approaches suffer from a lack of interpretability of the trained models. Here we present an interpretable machine learning framework for analyzing the nonlinear evolution dynamics of optical wavepackets in complex wave media. We use sparse regression to reduce microscopic discrete lattice models to simpler effective continuum models which can accurately describe the dynamics of the wavepacket envelope. 
We apply our approach to valley-Hall domain walls in honeycomb photonic lattices of laser-written waveguides with Kerr-type nonlinearity and different boundary shapes. 
The reconstructed equations accurately reproduce the linear dispersion and nonlinear effects including self-steepening and self-focusing. This scheme is proven free of the a priori limitations imposed by the underlying hierarchy of scales traditionally employed in asymptotic analytical methods. It represents a powerful interpretable machine learning technique of interest for advancing design capabilities in photonics and framing the complex interaction-driven dynamics in various topological materials.
\end{abstract}

\maketitle

\section{Introduction} 

Machine learning (ML), a branch of artificial intelligence, is revolutionizing various scientific fields by enabling pattern extraction and prediction from large datasets~\cite{RevModPhys.91.045002,chen2021physics,pilozzi2018machine,Yun_2022}. In recent years, the use of ML to determine the governing equations of various dynamical systems and processes has shown remarkable potential. 
One possible instrument for this purpose is the regression algorithm~\cite{Kaptanoglu2022, desilva2020}. 
In particular, this tool has been successfully tested on famous fundamental physical models described by partial differential equations (PDEs), such as nonlinear Burgers' and  Korteweg-de Vries equations~\cite{rudy2017data}, 
and the Belousov-Zhabotinsky reaction~\cite{rudy2017data, psindy1}. 
It was shown that the coefficients of governing PDEs of a known type can be recovered from numerical data generated by solving the equation within the same problem dimensionality.

The current challenge is to apply ML methods to real-world problems, moving beyond pre-expected basic models. Recently, beginning efforts in this direction have been reported. For instance, the authors of Ref.~\cite{naik2022discovering} used experimental data from degrading perovskite thin films subjected to environmental stressors to infer the underlying differential equation. Similarly, in nonlinear optics, data-driven approaches have been utilized to identify optimal conditions for four-wave mixing in an optical fiber~\cite{ermolaev2022data}.

\begin{figure*}[t!]
	\centering
\includegraphics[width=1\textwidth]{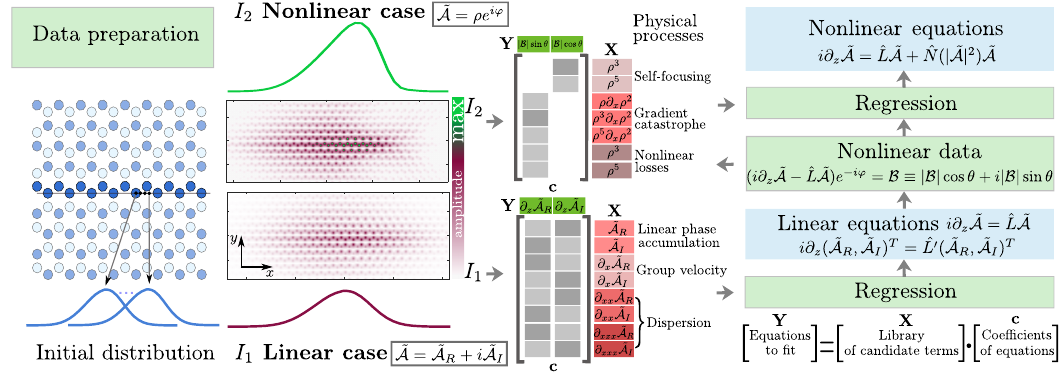}
\caption{
Schematic of the methodology for extracting the 
PDE model with ML using regression and numerically calculated datasets. 
The sequence of steps includes: data collection through paraxial modeling at low input powers, reconstruction of the linear operator terms, analysis of data calculated at higher input powers, and refinement of the PDE with the nonlinear terms.
}
\label{fig:2}
\end{figure*}

A particularly striking form of wave dynamics arises in topological materials, where complicated microscopic lattice models give rise to robust guided modes at edges or interfaces between distinct topological phases. Such edge-based form of transport has sparked significant interest in photonics, as it shows promise for constructing scattering-resistant transmission waveguiding channels in high-speed photonic circuits and communication networks~\cite{Ozawa2019}. 

The analysis of electromagnetic wave propagation along topological domain walls traditionally relies on numerical simulations. Analytical methods are based on simplifying assumptions to achieve a quasi-one-dimensional reduction in the long-wavelength limit from higher-dimensional PDEs. 
In the case of 2D lattices, this implies excluding the dimension transverse to the interface~\cite{kartashov2016modulational,PhysRevA.94.021801,smirnova2019topological,Kartashov2021}. One can additionally use multi-scale analysis to integrate out any microscopic spin-like degrees of freedom to obtain simpler scalar wave equations~\cite{smirnova2021gradient}.

More generally, dimensionality reduction is applicable to complex problems in other domains. 
For instance, the inherently nonlinear Navier-Stokes equations in hydrodynamics can be reduced to Lorentz-like ODEs to facilitate easier bifurcation analysis and identify conditions for chaotic dynamics corresponding to turbulence in the original system~\cite{Brunton2016}.
Another example is the nonlinear Schr\"odinger equation, analyzed using ODEs for moments -- integral characteristics of the beam. ML regression was recently tested to reconstruct the coefficients of these ODEs, instead of analytical means~\cite{Yang2024}.

However, the sophisticated geometry of topological lattices and the presence of optical nonlinearities often pose challenges in formulating reduced models that adequately capture the key properties needed to describe edge wavepacket dynamics. The derivations using asymptotic methods are cumbersome and closely tied to perturbation theory, addressing the hierarchy of characteristic scales~\cite{smirnova2021gradient}. From a ML perspective, this generally requires transitioning from multidimensional PDEs to simpler 1D PDEs, necessitating adaptation of the regression scheme.

In this work, we apply a data-driven ML for the first time in the context of complex nonlinear photonic lattices with nontrivial topology, demonstrating how ML regression can be used to obtain simpler yet accurate PDE models for the edge state dynamics. The usefulness of such a model lies in its predictive power, which comes from identifying underlying physical effects. As a specific example, we consider the valley-Hall domain wall created by inversion in feasible honeycomb staggered arrays of waveguides laser-written in glass, with parameters comparable to those used in the experiments of Ref.~\cite{Noh2018}. In this context, a nonlinear optical response arises from the intensity-dependent refractive index of the ambient glass medium.

We determine the PDE model governing the evolution of the edge wavepacket envelope as it propagates along the domain wall using a sparse-regression method based on data obtained from direct modeling. Our approach is summarized in Fig.~\ref{fig:2}. The lattice geometry in the $xy$ plane, transverse to the evolution direction $z$, is depicted on the left in Fig.~\ref{fig:2}. 
We seek a continuum model that is universally applicable to relatively broad wavepackets, with widths significantly exceeding the lattice period. 
The envelope function profile is extracted from the waveguides at the interface. In view of discreteness, 
more complete data are accumulated by sweeping the position of the input beam between the lattice sites. 

We use a split-step strategy to differentiate between linear and nonlinear scenarios and switch between them by tailoring the input power magnitude. Accordingly, for convenience, we represent the complex-valued function $\mathcal{A}$ either in terms of its real and imaginary parts or in terms of its intensity and phase. 
The regression is used to identify a subset of relevant terms from a large library of potential functions that best replicate the system's dynamics~\cite{Tibshirani1996}. After inputting $z$-series data, which is indicated in the Fig.~\ref{fig:2} as $\mathbf{Y}$, we construct a library of candidate functions (comprising the matrix $\mathbf{X}$)  and iteratively solve an optimization problem to obtain a sparse vector of coefficients $\mathbf{c}$ representing the unknown dynamical equations.

While the initial library contains an extended set of possible functions, the scheme allows filtering out absent or inessential contributions and retaining only the physical effects relevant for the studied  propagation distances.
The nonlinear terms in the PDE model at higher intensities are obtained as a refinement of the linear differential operator, initially reconstructed for low intensities in the linear regime.
This approach is somewhat analogous to the split-step solution for evolutionary problems and ML boosting, which iteratively improves the predictive power of the model by refining its weaker predictive versions~\cite{schapire2003boosting}.

\section{Data collection} \label{sec:phys_sys}
To model the paraxial evolution of light through the optical lattice along the $z$ axis (aligned parallel to the waveguides), we employ  the paraxial wave equation for the field $\mathcal{E}$:
\begin{multline} \label{eq:paraxial}
 i \frac{\partial \mathcal{E}}{\partial z}+\frac{1}{2 k_0} \Delta_{\perp} \mathcal{E}+\frac{k_0 n_L\left(\boldsymbol{r}_{\perp}\right)}{n_0} \mathcal{E} + n_2|\mathcal{E}|^2 \mathcal{E} = 0 \:, 
\end{multline}
where $\boldsymbol{r}_{\perp}=(x, y)$ are the transverse in-plane coordinates, $k_0=2 \pi n_0 / \lambda$ is the wave number, $n_0$ is the
background refractive index, 
$n_L\left(\boldsymbol{r}_{\perp}\right)$ is the perturbation of the refractive index forming the geometry of the lattice, $|n_L\left(\boldsymbol{r}_{\perp})\right|\ll n_0$. 
Eq.~\eqref{eq:paraxial} also includes a nonlinear term $\propto n_2|\mathcal{E}|^2 \mathcal{E}$, which is responsible for the focusing cubic Kerr-type nonlinearity.

The lattice geometry is imprinted into a refractive index distribution 
$n_L(x, y)\!=\! n_A \sum_{n, m} F\left(x-x_n, y-y_m\right)+n_B \sum_{n, m} F\left(x-x_n, y-y_m\right).$
Here $A$ and $B$ are the indices of the two triangular sublattices of the graphene lattice having spatial period $a$, and $n_A$ and $n_B$ refer to the perturbations of the refractive index in waveguides. The Gaussian-shaped elliptical waveguides have semiaxes $L_x$ and $L_y$, described by the function $F(x, y)=e^{-x^2 / L_x^2-y^2 / L_y^2}$.  The parameters corresponding to different degrees of parity breaking are listed in Table~\ref{Tab:paraxial}. 

\begin{table}[b!] 
\centering
\begin{tabular}{l|l|l} 
Parameter & Set I & Set II \\ \hline \hline
     $L_x$  &  $3.2 ~\mathrm{\mu m}$ & $4.9 ~\mathrm{\mu m}$\\
     $L_y$  &  $4.9 ~\mathrm{\mu m}$ & $3.2 ~\mathrm{\mu m}$\\
     $a_0$  &  $20 ~\mathrm{\mu m}$ & $18.5 ~\mathrm{\mu m}$\\
   $n_0$  &  $1.47$  &  $1.47$ \\
   $n_A$  &  $2.6 \times 10^{-3}$ & $7.5 \times 10^{-4}$  \\
   $n_B$  &  $2.8 \times 10^{-3}$ & $12.4 \times 10^{-4}$  \\
   $n_2$  &  $3 \times 10^{-20}~\mathrm{m^2/W}$  & $3 \times 10^{-20}~\mathrm{m^2/W}$\\
   $\lambda$ & $1650~\mathrm{nm}$ & $1045~\mathrm{nm}$\\
   $I_0$  &  $10^{16} ~\mathrm{W/m^2}$  & $10^{16} ~\mathrm{W/m^2}$
\end{tabular}
\caption{Two sets of lattice parameters simulated using paraxial modeling. }
\label{Tab:paraxial}
\end{table} 

\begin{figure}[t!]
	\centering
\includegraphics[width=8.4cm]{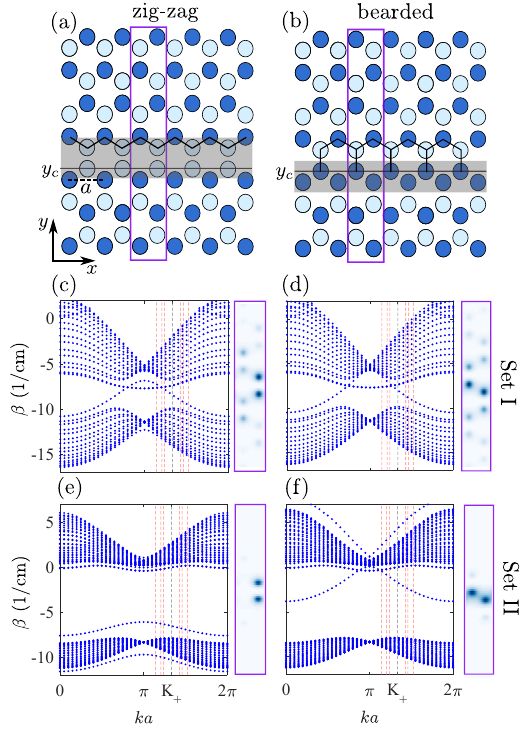}
\caption{
(a,b) Two distinct shapes of valley-Hall domain walls in a honeycomb lattice of laser-written waveguides: zigzag (a) and bearded (b). Parameter $a$ denotes the spatial lattice period. Supercells of these configurations are framed in violet rectangles, with the interface area of the domain wall shaded in gray. Panels (c-f) show the band structures, $\beta(k)$, for two parameter sets: Set I (upper row) and Set II (lower row). Calculations are performed for supercells of the staggered graphene lattice composed of 64 dielectric elliptical waveguides, incorporating domain walls of zigzag (c,e) and bearded (b,d) shapes, with open boundary conditions. Dashed vertical lines indicate the three scanning beam widths centered at wavenumber $K_+$.  
The corresponding transverse intensity profiles of the plane-wave-like edge state at the $K_+$ point are shown in the violet frames on the left.
} 
\label{fig:1}
\end{figure}

The interface between two domains is created by swapping the refractive index perturbations between two sublattices, $n_A \rightarrow n_B, n_B \rightarrow n_A$. This results in a domain wall where the refractive index perturbation is the same across the interfacing neighboring elements, as highlighted by the gray rectangle in Fig.~\ref{fig:1} (a,b).
In a dimerized honeycomb lattice, this interface can take two distinct shapes: zigzag and bearded shapes.

To prepare datasets, we solve the paraxial equation~\eqref{eq:paraxial} numerically. First, the plane wave expansion method is employed to obtain the edge mode profile transverse to the interface, $u(x,y)$, with the substitution $\mathcal{E}=u(x, y) e^{-i \beta z+i k x}$.
Then, the beam propagation method is used to simulate the evolution of the beams as they propagate along the waveguides' axis $z$. The beams are confined to the interface and localized along the interface with a Gaussian profile $f(x,{\bar{x}}_0)=f_0 e^{-(x-{\bar{x}}_0)^2/(2\mathcal{L}^2)}$ centered at ${\bar{x}}_0$.
Accordingly, the initial condition is set as $\mathcal{E}(z=0,x,y,{\bar{x}}_0)=f(x,{\bar{x}}_0) u(x,y) e^{i K_+ x}$, where $K_+=4\pi/(3a)$ is the high-symmetry point of the honeycomb lattice's Brillouin zone.  To capture general features, non-specific to initial conditions, and to deduce the corresponding effective model within reasonable propagation distances, simulations are performed for several different beam widths, $\mathcal{L}$. 

To describe the evolution in terms of the envelope function, we extract the values of the envelope at the centers of the waveguides forming the domain wall, 
$\bar{\mathcal{A}}(z,x_m,{\bar{x}}_0)=\mathcal{E}(z, x \equiv x_m,y \equiv y_c,{\bar{x}}_0)$, where $y_c$ is the vertical coordinate of the waveguides comprising the interface, see Fig.~\ref{fig:1}(a,b).
The sparse nature of the function $\bar{\mathcal{A}}(z,x_m,{\bar{x}}_0)$ along the $x$-axis, defined within the waveguides at points $x=x_m$, presents a potential challenge in accurately determining the governing continuum equation, where fine discretization is typically assumed. The latter is particularly important for providing an accurate approximation of derivatives in the PDE model. To circumvent this challenge, we set a range of initial conditions and calculate multiple envelopes to 
generate more complete data.
Specifically, we mesh the intervals 
$[x_m,x_{m+1}]=[x_m,x_m+\Delta_x q]$, where the number of steps $q=32$ and step size $\Delta_x = a/q $, and sweep ${\bar{x}}_0$ such that ${\bar{x}}_0(n)={\bar{x}}_0+\Delta_x n$, $n=[1,...,q]$. This enables us to perform several calculations of $\bar{\mathcal{A}}(z,x_m,{\bar{x}}_0(n))$, 
which can then be 
combined into a smooth function $\mathcal{A}(z,x)$. 

\begin{figure}[b!]
	\centering
\includegraphics[width=8.4cm]{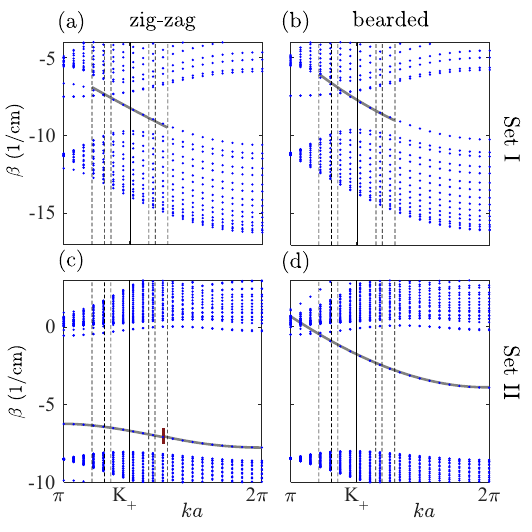}
\caption{ Comparison between the numerical (blue dots) edge-state dispersion curve in the band structures and the dispersion plotted using ML-determined coefficients (black lines). Parameters for panels (a-d) correspond to Figs.~\ref{fig:1}(c-f), respectively. 
Panels (a), (b), and (d) show the ML-based approximation recovered near $K_+$. In plot (c), the approximation to the left and to the right of the brown vertical line segment are recovered near $ K_+$ and $2\pi$, respectively.
}
\label{fig:d}
\end{figure}

In Ref.~\cite{smirnova2021gradient}, we used an analytical asymptotic procedure  to derive the evolution equation 
for the slowly varying amplitude $\tilde{\mathcal{A}}(z,x)$ of edge wave packets in the continuum limit, applicable regardless of the shape of the domain wall, to both zigzag and bearded cuts,
\begin{equation}
\label{eq:solw}
i\frac{\partial \tilde{\mathcal{A}}}{\partial {z}} =  i v \frac{\partial \tilde{\mathcal{A}}}{\partial {x}} - G |\tilde{\mathcal{A}}|^2 \tilde{\mathcal{A}} - i v_g {|\tilde{\mathcal{A}}|^2 \dfrac{\partial |\tilde{\mathcal{A}}|^2} {\partial x} \tilde{\mathcal{A}}} - \eta \dfrac{\partial^2 \tilde{\mathcal{A}}}{\partial x ^2} + \beta_0 \tilde{\mathcal{A}}
 \:.
\end{equation} 
However, the asymptotic scope is always limited by a priori assumptions for the lattice parameters, which narrow the range of applicability to specific conditions.
Instead of performing series expansions at different orders, the aim of the present study is to determine the structure of the equation and its coefficients using ML techniques and data for the numerically extracted envelope ${\mathcal{A}}(z,x)$.
For this, we developed custom code implementing the linear regression algorithm.

We select a dataset consisting of randomly mixed points $(z_j,x_j)$ and compute the matrix $\mathbf{X}$ by utilizing ${\mathcal{A}}(z_j,x_j)$ and its numerically obtained derivatives.
When collecting data, we use only the points where the intensity exceeds some threshold. For our purposes, it is essential to focus on and accurately describe the dynamics of the field within the beam localization region. Including all data points, especially those with very low values, can introduce uncertainty and ambiguity in phase calculations and lead to inaccurate derivative computations. 
Randomly shuffled data points are divided into 80\% for training and 20\% for testing. 
Coefficients are determined from the larger training set, and then these coefficients are tested on the test set. 
Additionally, the coefficients are validated by comparing numerical solutions of the full paraxial and the PDE models~\cite{Suppl}.

\begin{figure}[t!]
	\centering
\includegraphics[width=8.4cm]{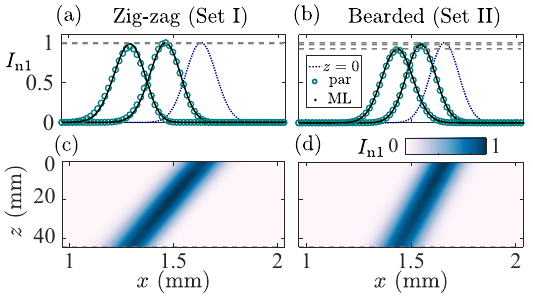}
\caption{ 
Evolution of the beam envelope at zigzag (left column) and bearded (right column) domain walls at low input intensity, $I(z=0)<0.04 \times I_0$. 
Plotted is the intensity, normalized to its maximum, $I_{\text{n}1}$. 
Panels (a,b) show snapshots taken at three propagation distances: at the input ($z=0$, blue dotted line), at $z=22~\mathrm{mm}$, and at $z=44~\mathrm{mm}$. Green circles: paraxial modeling results; black solid lines: numerical solution of PDE with ML-determined coefficients.
(c,d) Spatial profile mapped along the $z$-directed propagation.
}
\label{fig:3}
\end{figure}

\section{Linear low-intensity regime}
\label{sec:lin}
We begin with the low-intensity linear case.
The envelope equation can be formulated as a system for the real and imaginary parts in the representation $\tilde{\mathcal{A}}=\tilde{\mathcal{A}}_R+i\tilde{\mathcal{A}}_I$: 
\begin{subequations}
\begin{multline}
\frac{\partial\tilde{\mathcal{A}}_R}{\partial z}= \beta_0 \tilde{\mathcal{A}}_I + \beta_{0I} \tilde{\mathcal{A}}_R + v\frac{\partial \tilde{\mathcal{A}}_R}{\partial x} - v_I \frac{\partial \tilde{\mathcal{A}}_I}{\partial x} \\
- \eta \frac{\partial^2 \tilde{\mathcal{A}}_I}{\partial x^2} - \eta_I \frac{\partial^2 \tilde{\mathcal{A}}_R}{\partial x^2} + \eta' \frac{\partial^3 \tilde{\mathcal{A}}_I}{\partial x^3} + \eta'_I \frac{\partial^3 \tilde{\mathcal{A}}_R}{\partial x^3}\:,
\end{multline}
\begin{multline}
  \frac{\partial \tilde{\mathcal{A}}_I}{\partial z}= - \beta_0 \tilde{\mathcal{A}}_R + \beta_{0I} \tilde{\mathcal{A}}_I  + v\frac{\partial \tilde{\mathcal{A}}_I}{\partial x} + v_I \frac{\partial \tilde{\mathcal{A}}_R}{\partial x} \\
 + \eta \frac{\partial^2 \tilde{\mathcal{A}}_R}{\partial x^2} -  \eta_I \frac{\partial^2 \tilde{\mathcal{A}}_I}{\partial x^2}- \eta' \frac{\partial^3 \tilde{\mathcal{A}}_R}{\partial x^3} + \eta'_I \frac{\partial^3 \tilde{\mathcal{A}}_I}{\partial x^3}.
\end{multline}
\end{subequations}
The library of functions consists of differential operators of various orders for recovering spatial dispersion. This essentially addresses the problem of reconstructing the dispersion of the edge state by scanning it with a beam of finite spectral width in the vicinity of the specified wave vector, see Fig.~\ref{fig:d}. 
The real and imaginary parts representation is optimal for this problem, as it enables a more compact function library for differential operators compared to the intensity-phase representation.

The schematic organization of the library is partially visualized in Fig.~\ref{fig:2}. Even when we assume all the coefficients in the equations for $\tilde{\mathcal{A}}_R$ and $\tilde{\mathcal{A}}_I$  are different, some of these coefficients 
group according to physical processes, while others are minor and can be omitted, following the principles of sparse regression~\cite{Suppl}.
The coefficients of the dominant terms are then further refined.
The leading terms at short propagation distances stem from linear phase accumulation, drift with group velocity, and second-order dispersion, which causes gradual symmetric broadening.
These are described by the coefficients 
$\langle \mathbf{c}_i \rangle = \langle \beta_0 \rangle, \langle v \rangle, \langle \eta \rangle$ and their standard deviations
$\delta \mathbf{c}_i$
calculated on 100 validation folds of the train dataset~\cite{Suppl}.
The regimes of undistorted propagation and broadening, accompanied by a decrease in amplitude, are illustrated in Fig.~\ref{fig:3}.  
At longer propagation distances, the model can be further refined by incorporating third-derivative terms responsible for the asymmetric distortions.

\section{ Nonlinear high-intensity regime}
\label{sec:nonl}

\begin{figure}[t!]
	\centering
\includegraphics[width=8.4cm]{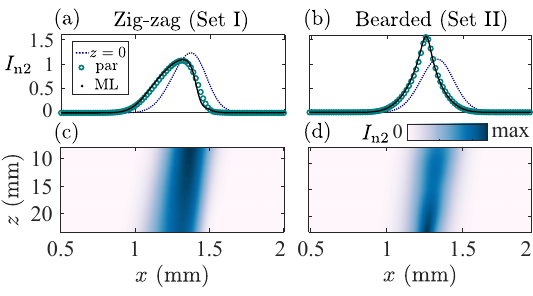}
\caption{ 
The nonlinear scenarios of beam propagation at
 zigzag (left column) and bearded (right column) domain walls. 
 Similar to Fig.~\ref{fig:3}, panels (a,b) show snapshots taken at the input ($z=7.5~\mathrm{mm}$, blue dotted line), and at $z=23~\mathrm{mm}$. 
The initial intensity is 
 ${I}(z=0)=(0.425)^2\times I_0$, plotted is $I_{\text{n}2} ={I}(z)/[(0.4)^2 \times I_0]$. (c,d) Nonlinear beam transformation mapped along $z$.
}
\label{fig:4}
\end{figure}

As the optical power increases, we anticipate the manifestations of nonlinear effects such as self-focusing or self-steepening originating from the nonlinear group velocity. Therefore, it is natural to focus on these effects and evaluate whether they suffice to accurately describe the beam's transformation at certain distances.
The library of functions is assembled to fit the following equation
\begin{multline}
i\frac{\partial \tilde{\mathcal{A}}}{\partial {z}} =  i v \frac{\partial \tilde{\mathcal{A}}}{\partial {x}} - \eta \dfrac{\partial^2 \tilde{\mathcal{A}}}{\partial x ^2} + i \eta'_I \dfrac{\partial^3  \tilde{\mathcal{A}}}{\partial x ^3} + \beta_0 \tilde{\mathcal{A}} \\
  - i (v_{g1} \tilde{\mathcal{A}}+v_{g2} {|\tilde{\mathcal{A}}|^2 \tilde{\mathcal{A}}}+v_{g3} {|\tilde{\mathcal{A}}|^4 \tilde{\mathcal{A}}} )\dfrac{\partial |\tilde{\mathcal{A}}|^2} {\partial x}-\\
   - (G_1 |\tilde{\mathcal{A}}|^2+G_2 |\tilde{\mathcal{A}}|^4   )  \tilde{\mathcal{A}} + i (\gamma_1 |\tilde{\mathcal{A}}|^2+\gamma_2 |\tilde{\mathcal{A}}|^4)\tilde{\mathcal{A}}\:.
\end{multline}
Within the intensity-phase representation, $\tilde{\mathcal{A}} = \rho e^{i \varphi}$, it is rewritten as a system of equations for $\rho$ and $\varphi$. 
Remarkably, nonlinear effects appear separated in this framework: self-focusing contributes to the phase, while loss and self-steepening effects are incorporated into the intensity equation. 
We assume that the coefficients of the linear operator have already been determined from the low-intensity analysis. Therefore, we now solve the regression problem for the difference between the evolution operator and the linear operator, multiplied by the phase, $(i\partial_z \tilde{\mathcal{A}} - \hat{L} \tilde{\mathcal{A}})e^{-i\varphi} \equiv |\mathcal{B}| \cos \theta + i |\mathcal{B}| \sin \theta $, which can be attributed to the nonlinear correction $\hat{N} \tilde{\mathcal{A}}$ showing up at higher intensities, 
\begin{subequations}
\begin{equation}
- |\mathcal{B}| \cos \theta =-\beta_0 \rho
+G_1 \rho^3+G_2 \rho^5  \:,
\end{equation}
\begin{equation}
 |\mathcal{B}| \sin \theta = - (v_{g1} \rho+ v_{g2} \rho^3 +v_{g3} \rho^5  ) \frac{\partial \rho^2}{\partial x}+\gamma_1 \rho^3 +\gamma_2 \rho^5
 \:.
 \end{equation}
\end{subequations}

The determined coefficients confirm the presence of two competing major nonlinear effects. Figure~\ref{fig:4} shows representative cases of the nonlinear dynamics. A growing asymmetry of the wavepacket is visible in Fig.~\ref{fig:4}(a) and (c). This self-steepening deformation occurs due to the prevailing nonlinear velocity term $v_g$~\cite{smirnova2021gradient,Smolina2023}. 
In contrast, gradual self-compression, a hallmark of self-focusing~\cite{kartashov2016modulational,Kartashov2021,smirnova2021gradient}, is evident in Fig.~\ref{fig:4}(b) and (d). The latter behavior is typical of the nonlinear Schr\"odinger equation with attractive nonlinearity and dominant quadratic spatial dispersion.

Which nonlinear effect is dominant depends on the microscopic lattice parameters. Conventionally, the corresponding nonlinear coefficients would be computed in terms of integrals of the edge states' spatial profiles transverse to the interface when applying a conventional asymptotic multi-scale analysis~\cite{smirnova2019topological,smirnova2021gradient}. The present data-driven approach allows one to bypass this step and infer the corresponding nonlinear coefficients purely using beam propagation simulations.

\section{Conclusion}

In this study, we demonstrated how ML can effectively unveil the continuum PDE model governing edge waves confined to domain walls in photonic lattices.
It enables the revelation of various effects, including spatial dispersion and higher-order nonlinearities, and the resultant corrections to the nonlinear Schr\"odinger equation, such as the nonlinear velocity term, for domain walls of different cuts.
Thus, our approach provides a valuable alternative to traditional fully analytical asymptotic methods for addressing wave dynamics in nonlinear and crystalline systems, allowing for the exploration of distinct topological phases and applicable beyond photonics to various wave media. It can readily be extended to other types of nonlinearities and interparticle interactions, for example in polaritonic platforms.


\end{document}